\documentclass[]{osa-article}
\journal{oe}
\articletype{Research Article}

\begin{document}

\title{Nanostructure-modulated planar high spectral resolution spectro-polarimeter}

\author{L. Pjotr Stoevelaar,\authormark{1,2,+,*}
Jonas Berzin\v{s},\authormark{1,3,+,*}
Fabrizio Silvestri,\authormark{1}
Stefan Fasold,\authormark{3}
Khosro Zangeneh Kamali,\authormark{4}
Heiko Knopf,\authormark{3,5,6}
Falk Eilenberger,\authormark{3,5,6}
Frank Setzpfandt,\authormark{3}
Thomas Pertsch,\authormark{3,5}
Stefan M.B. B{\"a}umer,\authormark{1}
and Giampiero Gerini\authormark{1,2}}

\address{\authormark{1}The Netherlands Organization for Applied Scientific Research, TNO, Optics Department, 2628CK Delft, The Netherlands\\
\authormark{2}Eindhoven University of Technology, TU/e, Electromagnetics Group, 5600MB Eindhoven, The Netherlands\\
\authormark{3}Institute of Applied Physics, Abbe Center of Photonics, Friedrich Schiller University Jena, 07745 Jena, Germany\\
\authormark{4}Nonlinear Physics Centre, The Australian National University, Canberra ACT 2601, Australia\\
\authormark{5}Fraunhofer Institute for Applied Optics and Precision Engineering, 07745 Jena, Germany\\
\authormark{6}Max Planck School of Photonics, 07745 Jena, Germany\\
\authormark{+}These authors contributed equally to this paper.\\
}

\email{\authormark{*}pjotr.stoevelaar@tno.nl, jonas.berzins@uni-jena.de} 
% Frank suggested to put something like "These authors have contributed equally." Will look for example.

%%%%%%%%%%%%%%%%%%%%%%%%%%  abstract  %%%%%%%%%%%%%%%%%%%%%%%%%%
\begin{abstract}
We present a planar spectro-polarimeter based on Fabry-P{\'e}rot cavities with embedded polarization-sensitive high-index nanostructures. A $7~\upmu$m-thick spectro-polarimetric system for 3 spectral bands and 2 linear polarization states is experimentally demonstrated. Furthermore, an optimal design is theoretically proposed, estimating that a system with a bandwidth of 127~nm and a spectral resolution of 1~nm is able to reconstruct the first three Stokes parameters \textcolor{black}{with a signal-to-noise ratio of -13.14~dB with respect to the the shot noise limited SNR}. The pixelated spectro-polarimetric system can be directly integrated on a sensor, thus enabling applicability in a variety of miniaturized optical devices, including but not limited to satellites for Earth observation. 
\end{abstract}

%%%%%%%%%%%%%%%%%%%%%%%%%%  body  %%%%%%%%%%%%%%%%%%%%%%%%%%
\section{Introduction}
Multifunctional imaging has emerged as a new generation of digital imaging. Techniques such as polarimetry and hyperspectral imaging provide substantially more information on the imaged scene or the object of interest than the conventional color imaging~\cite{Snik2014,Khan2018,hillier1999}. Information on the polarization state of the collected light enables a better understanding of surface topography and scattering, thus is widely used for target detection in defense and biomedical applications~\cite{Snik2014}, while a high-resolution spectral information provides details on the material composition for the assessment of food quality, artwork authentication and many other applications~\cite{Khan2018}. All these techniques fall under the term of spectro-polarimetry~\cite{hillier1999}. 

One of the main applications of spectro-polarimetry is Earth observation~\cite{Snik2014}, where the properties of aerosols, e.g. their size, shape, and refractive index, can be identified remotely using polarization and spectral information. In general, astronomical instruments on board of satellites have stringent constraints in terms of mass and volume. Therefore, any innovative solutions that enable compact instruments is highly desirable, especially with the increased number of nano- and cube-satellites in recent years ~\cite{cubesat2019}. Accordingly, there is an increasing demand for optical components to be miniaturized. However, despite many efforts,  current spectro-polarimetric systems in space still consist of multiple thick optical elements and result in cumbersome payloads~\cite{Deschamps1994,VanHarten2011}. 

The size limitation of the conventional optical components motivated the search for alternative implementations, where the concept of metasurfaces has emerged as one of the most promising technologies. A metasurface is a two-dimensional nanostructure array, which enables control of amplitude, phase, and polarization of the incident light~\cite{Arbabi2015}. It has been successfully used in the realizations of polarimeters~\cite{Yue2016,Wei2017,Rubin2018, arbabi2018, yan2019}, spectral filters and spectrometers ~\cite{Faraji-Dana2018,Horie2016,Yue2015a,Yue2015,Shaltout2018,Berzins2019}, and even spectro-polarimeters~\cite{Chen2016a,ding2017,Tu2018,li2019}. However, there are still several obstacles to overcome and challenges to be addressed. First, the majority of multifunctional nanostructure-based devices consist of a few transversely-variant layers on top of each other. This poses quite a challenge in terms of alignment, as even small misalignment causes errors in the detected information~\cite{Tu2018}. Furthermore, for the maximum gain in the spectral information, it is important to obtain a high spectral resolution, but up to now, the spectral resolution of spectro-polarimetric systems has been limited to the bands of RGB filters~\cite{Tu2018}, or required a connection to an external spectrometer~\cite{Chen2016a}. Last but not least, imaging systems, ideally, should be integrated on a compact sensor, thus aspects as compatibility, thickness and height uniformity of the pixels are of high importance and have to be addressed. 

In this work, we introduce a planar spectro-polarimeter concept, which is based on a set of polarization-sensitive silicon (Si) nanostructures embedded in a Fabry-P{\'e}rot (FP) cavity and which could be directly integrated on a sensor~\cite{TNOpatent}. Such a system enables the reconstruction of the first three Stokes parameters ($S_0$, $S_1$, $S_2$) with a high spectral resolution. In this paper, we present its systematic design, experimental demonstration, and its comparison to an ideal case. Furthermore, we discuss the polarization reconstruction algorithm, its limitations, and optimization towards multi-band spectro-polarimetric designs. 

\section{Methods}
%%%%%%%%%%%%%%%%%%%%%%%%%%%%%%%%%%%%%%%%%%%%%%%%%%%%%%%%%%%%%%%%
\subsection{Concept} \label{concept}
\begin{figure}[t]
    \centering
    \includegraphics[width=1\linewidth]{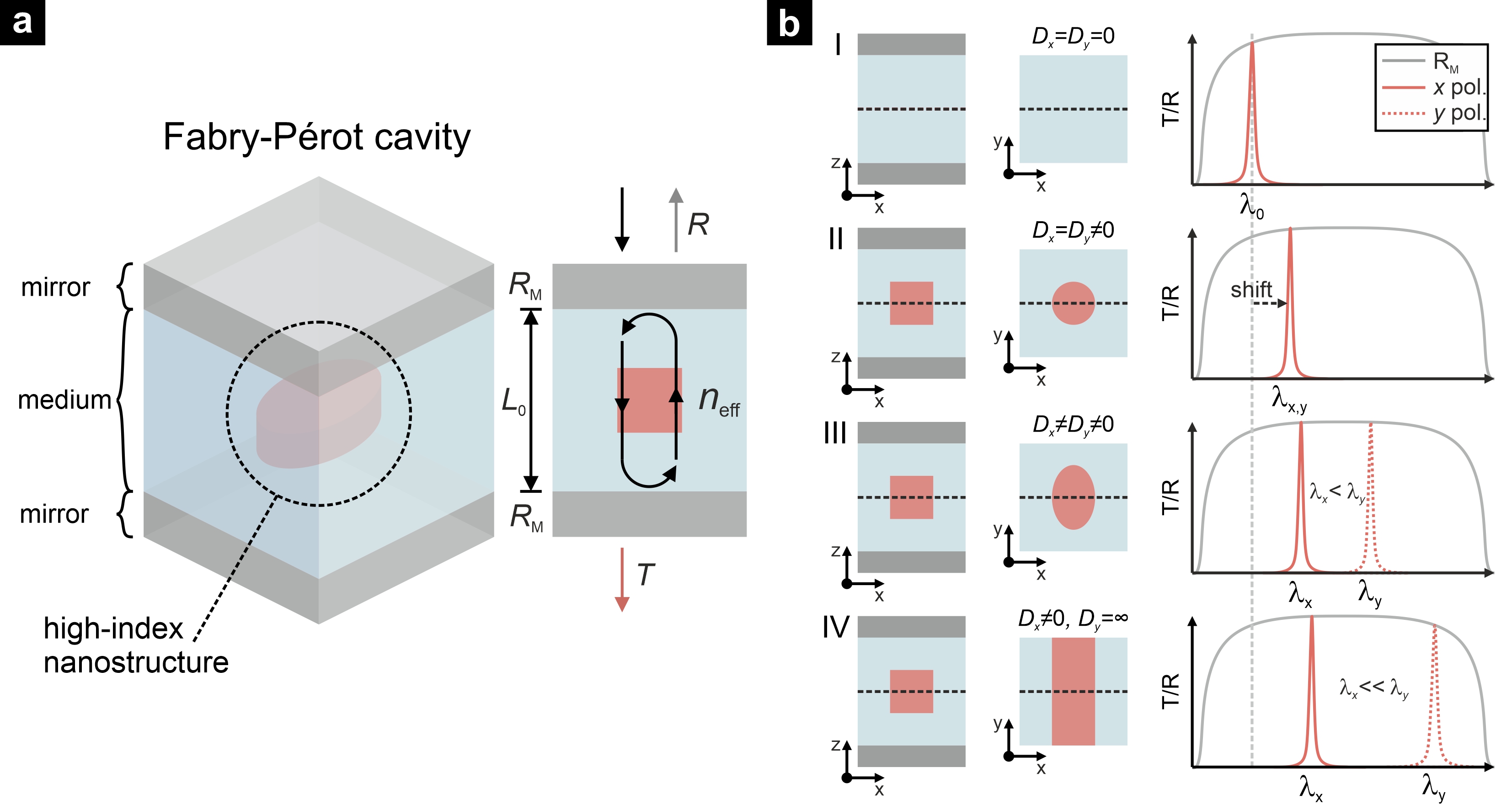}
    \caption{Concept of FP cavity modulated by high-index nanostructures. (a)~Illustration of FP cavity constituted of mirrors with reflectivity $R_\text{m}=R_1=R_2$ spaced by medium with optical length $L=L_0\cdot n_{\text{eff}}$. Cavity is modulated by inclusion - high-index nanostructure. (b)~Regimes of FP cavity depending on the shape of the nanostructure: I)~empty cavity (without inclusion) producing a single peak at in the spectral range of interest; II)~cavity modulated by polarization-insensitive nanostructure, single peak remains but is red-shifted; III)~cavity modulated by polarization-insensitive nanostructure, two different peaks at $\lambda_x$ and $\lambda_y$ ($\lambda_x < \lambda_y$) for two linear polarization states, $x$ and $y$, respectively; IV)~cavity modulated by grating, two different peaks at $\lambda_x$ and $\lambda_y$ (but $\lambda_x \gg \lambda_y$) for two linear polarization states, $x$ and $y$, respectively. Transmittance of $x$-polarized light is depicted by red solid line, transmittance of $y$-polarized -- red dashed line, reflectivity band of the mirrors -- grey line.  
    }\label{fig1}
\end{figure}

A FP cavity consists of two mirrors separated by an optical length $L$, as shown in Fig.~\ref{fig1}(a). For simplicity, we assume that both mirrors are planar, their reflectivities are equal $R_\text{m}=R_1=R_2$, and the losses generated by scattering and absorption are negligible. The FP cavity provides a  Lorentzian-shape peak in transmission centered, at the wavelengths $\lambda_\text{c}$, for which the following condition holds~\cite{hodgson2005}:
\begin{equation}
\lambda_\text{c} = \frac{2L}{q},
\end{equation} 
where $q$ is an integral number, representing the order of the resonant mode. Ideally, the transmission exhibits a periodic sequence of peaks, but may also be limited to one by the reflectivity band of the mirrors, as illustrated in Fig.~\ref{fig1}(b). The central wavelengths of the peaks $\lambda_\text{c}$ can be tailored by the optical distance of the mirrors $L=L_0\cdot n_{\text{eff}}$, which rely on the geometrical path length between the mirrors $L_0$ and the effective refractive index of the cavity medium $n_{\text{eff}}$. 

The most common approach to tailor the transmission is to change the physical length of the cavity $L_0$~\cite{Wang2018}. However, it has been recently demonstrated that inclusions, such as arrays of high-index nanostructures, might be used to modulate the effective index $n_{\text{eff}}$~\cite{Horie2016}. The inclusion of high-index nanostructures adds a phase shift to the cavity and a subsequent red-shift of the resonant peak, see Fig.~\ref{fig1}(b). Furthermore, if the diameter in one axis $D_{x}$ of the nanostructures is chosen differently than the diameter of other $D_{y}$, the cavity will produce two different transmission peaks for two different linear polarizations: one corresponding to the case where the electric-field of the incident light aligns with the width of the nanostructure ($x$-polarized), the other - when the field aligns with the length of the nanostructure ($y$-polarized)~\cite{TNOpatent}, as shown in Fig.~\ref{fig1}(b). The spectral distance between the peaks can be controlled by altering the aspect ratio ($D_x/D_y$) of the nanostructures, while the largest spectral separation of the peaks can be achieved with a linear grating, since it has an infinite aspect ratio, see Appendix. In any case, it is important that the optical size of the nanostructures is significantly smaller than the wavelength of operation ($nD_{x,y}<\lambda$). In fact in this case, the nanostructures are non-resonant and the modulation of the cavity is based only on the change of the effective index~\cite{choy2015}.

%Such setup provides up to twice as much light compared to a conventional filter and polarizer setup, as a single pixel utilizes two peaks for the two different polarizations.

\begin{figure}[t]
   \centering
   \includegraphics[width=1\linewidth]{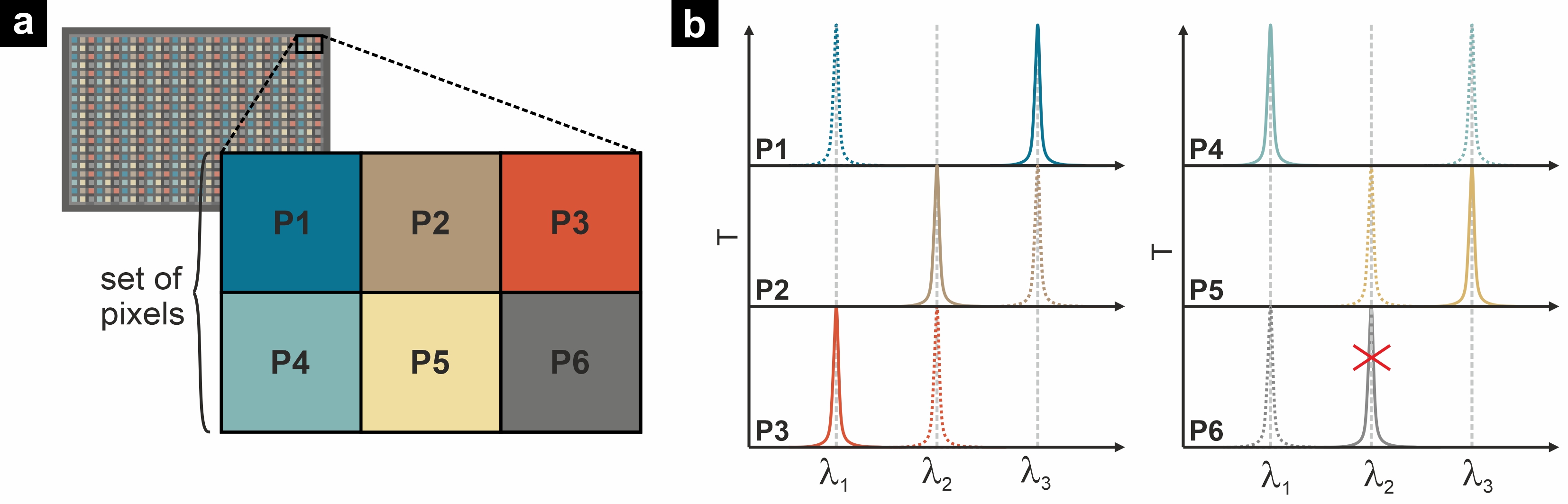}
   \caption{Concept of spectro-polarimeter for linear-polarization. (a) Illustration of a set of 6 pixels, a super-pixel, integrated directly on a sensor. (b) Exemplary Lorentzian-shape spectral functions for 3 wavelengths ($\lambda_1$, $\lambda_2$, $\lambda_3$) and two polarization states ($x$ and $y$) using set of 6 pixels ($\text{P1},\cdots,\text{P6}$). A single pixel provides two peaks for the two linear-polarizations, while one peak ($\text{P6}$) is cut-out for inversion of the matrix. Transmittance of $x$-polarized light is depicted by solid line, while transmittance of $y$-polarized light is shown in dashed line. 
}\label{fig2}
\end{figure}

Using different-size polarization-sensitive nanostructures, resonance peaks for different wavelength and polarization combinations can be obtained. By measuring a set of 6 pixel intensities ($I_1,\cdots,I_6$), illustrated in Fig.~\ref{fig2}(a,b), we can retrieve both spectral and polarization information in the form of discrete intensities of two linear polarization states ($I_x$ and $I_y$) at three wavelengths \{$\lambda_1,\lambda_2,\lambda_3$\}~\cite{TNOpatent}. For simplicity, it is assumed that the pixels have a transmission $\text{T} = \alpha_m^n$ at their peaks, where $n$ is the number corresponding to the detector pixel and $m$ indicates the polarization state and $\text{T} = 0$ elsewhere. By using this assumption, the following system of equations is formed:
\begin{equation}
    \left(\begin{array}{c}
         I_1\\
         I_2\\
         I_3\\
         I_4\\
         I_5\\
         I_6
    \end{array}
    \right)=\left[\begin{array}{cccccc}
         0             &0             &\alpha^{1}_{x}&\alpha^{1}_{y}&0             &0             \\
         0             &\alpha^{2}_{x}&0             &0             &0             &\alpha^{2}_{y}\\
         \alpha^{3}_{x}&0             &0             &0             &\alpha^{3}_{y}&0             \\
         \alpha^{4}_{x}&0             &0             &0             &0             &\alpha^{4}_{y}\\
         0             &0             &\alpha^{5}_{x}&0             &\alpha^{5}_{y}&0             \\
         0             &\alpha^{6}_{x}&0             &\alpha^{6}_{y}&0             &0
    \end{array}\right]
        \left(\begin{array}{c}
         I_x^{\lambda_1}\\
         I_x^{\lambda_2}\\
         I_x^{\lambda_3}\\
         I_y^{\lambda_1}\\
         I_y^{\lambda_2}\\
         I_y^{\lambda_3}
    \end{array}\right).
    \label{inversionmatrix}
\end{equation}
Since each of the measured values (intensities) is a combination of two unknowns, this system of equations is not full rank. Therefore, it cannot be inverted, making the spectro-polarimetric reconstruction impossible. However, it can be made invertable by removing at least one of the transmission peaks from the system of equations, e.g. by setting $\alpha^{6}_{x}=0$. The resulting matrix will be full rank and, subsequently, can be inverted making the spectro-polarimetric reconstruction possible. This results in the set of spectro-polarimetric pixels shown in Fig.~\ref{fig2}(a) with their spectral functions illustrated in Fig.~\ref{fig2}(b).

Furthermore, by including a second set of 6 pixels with the nanostructures rotated by an angle of $45^\circ$, the intensity of light polarized along angles of $45^\circ$ and $135^\circ$ can also be measured. This allows the retrieval of the first three Stokes parameters for different wavelengths:
\begin{align}
    \label{S0}
    S_0&=I_x+I_y=I_{45^\circ}+I_{135^\circ},\\
    \label{S1}
     S_1&=I_x-I_y,\\
    \label{S2}
    S_2&=I_{45^\circ}-I_{135^\circ}.
\end{align}
The system is not able to retrieve all Stokes parameters~\cite{Azzam2016}, as it is unable to distinguish circularly polarized light from unpolarized light (the fourth Stokes parameter ($S_3$) can not be retrieved), and the presence of circularly polarized light may disrupt the estimation of the first Stokes parameter. However, for some applications this is not critical, e.g. in Earth observation, where the amount of circularly polarized light is negligible.

%%%%%%%%%%%%%%%%%%%%%%%%%%%%%%%%%%%%%%%%%%%%%%%%%%%%%%%%%%%%%%%%
\subsection{Design} \label{design}
As the spectro-polarimeter is based on nanostructure-modulated FP cavities, its design starts from the mirrors. An important aspect of the FP cavity is its spectral full width at half maximum (FWHM), which relates to the spectral resolution of the system, see Section \ref{discussion}. Following the assumption that both mirrors have the same reflectivity $R_{\text{m}}$, the FWHM is given by the following equation~\cite{atherton1981}:
\begin{equation}
    \text{FWHM}=\frac{\lambda (1-R_{\text{m}})}{q \pi \sqrt{R_{\text{m}}}} ,\\
\end{equation}
thus the higher the reflectivity of the mirrors, the higher the spectral resolution. Because of that, distributed Bragg reflectors (DBRs) are used. If the DBR layers are made of lossless materials, there is no strict limitation to the number of layers and, subsequently, the FWHM~\cite{Knopf2019}. However, we restrained ourselves to DBRs of 7 pairs of alternating SiO$_{2}$ ($n = 1.45$) and TiO$_{2}$ ($n = 2.285$ ~\cite{siefke2016}) layers, with reflection band centered at $\lambda=1500$~nm.  This configuration resulted in reflectance $R_{\text{m}}\approx99.5\%$ and $\text{FWHM}\approx1$~nm (cavity quality factor $Q > 1000$). The mirrors were separated by an optical length $L$ equal to the central wavelength $\lambda=1500$~nm. The physical length of the cavity was set to $L_0=985$~nm. 

The arrays of polarization-sensitive Si nanostructures were made from amorphous hydrogenated Si (a-Si:H), due to its high refractive index and no intrinsic losses ($n = 3.7$ and $k = 0$ at $\lambda = 1500$~nm). The nanostructures were defined as extruded ellipsis with a certain diameter in $x$-axis ($D_{x}$), a diameter in $y$-axis ($D_{y}$), and a height $H$. The nanostructures were distributed in a square lattice with a period $P = 500$~nm. The height of the nanostructures was set to $H=300$~nm. The diameters of the elliptical nanostructures, $D_{x}$ and $D_{y}$, were optimized in the range 140-360~nm to obtain three spectral bands centered at the following wavelengths: $\lambda_1=1480$~nm, $\lambda_2=1500$~nm, and $\lambda_3=1520$~nm. The optimized nanostructures were placed in the middle of the cavity in order not to alter the performance of the mirrors. $\text{SiO}_{2}$ ($n = 1.45$) was selected as a low-index material of the cavity medium. Moreover, to remove a transmission peak in the matrix, as mentioned in Sec.~\ref{concept} and shown in Fig.~\ref{fig2}(b), one of the pixels with elliptical Si nanostructures was replaced by a subdiffractive Si grating. The grating was defined with an infinite aspect ratio, meaning a certain width $D_{x}$, but $D_{y}=\infty$. Regardless of the shape of the nanostructures inside the cavity, their optical size has restriction to be relatively smaller than $\lambda = 1500$~nm for the nanostructures to be non-resonant and provide transmission close to unity.

The design spectral range and the FWHM of the peaks was based on the extended requirements of the spectropolarimeter for planetary exploration (SPEX) ~\cite{van2017spex}. The design was obtained using a finite-difference time-domain (FDTD) method (Lumerical, Inc.).

%%%%%%%%%%%%%%%%%%%%%%%%%%%%%%%%%%%%%%%%%%%%%%%%%%%%%%%%%%%%%%%%
\subsection{Fabrication}\label{fabrication}
The sample fabrication was carried out in several steps: the deposition of the bottom mirror, the structuring of the cavity, and the deposition of the top mirror. 

First, the bottom mirror was based on alternating SiO$_2$ and TiO$_2$ layers. The layers were deposited on top of a glass substrate by plasma-ion-assisted deposition (PIAD), described in a previous publication~\cite{Knopf2019}. In total, seven pairs of SiO$_2$/TiO$_2$ layes were deposited for high reflectivity ($R_{\text{m}} \approx 98\%$) at $\lambda = 1500$~nm, see Appendix. The thicknesses of the layers were initially calculated as quarter-wavelength layers, but later tuned to $t_{\text{1L}} = 320$~nm and $t_{\text{1H}} = 120$~nm for SiO$_2$ and TiO$_2$, respectively, to reduce the thickness of TiO$_2$ layers in attempt to avoid its polycrystalline growth, thus retaining smooth surfaces with low scattering and absorbance~\cite{bennett1989}.

The nanostructures were made from a 300~nm layer of a-Si:H, deposited by a plasma-enhanced chemical vapor deposition (PECVD) using $\text{SiH}_4$ gas as a source (Oxford Plasmalab 100 Dual Frequency, Oxford Instruments). Afterwards, a 30 nm chromium (Cr) layer was deposited by ion beam deposition (Oxford Ionfab 300, Oxford Instruments) and a 100 nm layer of electron beam resist (EN038, Tokyo Ohka Kogyo Co., Ltd.) was spin-coated on top. Such sample was exposed by a variable-shaped electron-beam lithography system (Vistec SB 350, Vistec Electron Beam GmbH). First, the resist was developed and the mask was transferred in the Cr layer by ion beam etching (Oxford Ionfab 300, Oxford Instruments). Then, the Cr mask was transferred in the Si layer by inductively coupled plasma reactive ion etching (SI-500 C, Sentech Instruments GmbH) with $\text{CF}_4$ as reactive gas. Finally, the remaining resist and Cr mask was supposed to be removed by acetone and Cr etchant, but during the etching some of Cr mixed with other materials and was not fully removed. Cr has a high absorption ($n = 3.675$ and $k = 4.072$ at $1500$~nm ~\cite{johnson1974}). Any amount of it in the cavity is undesirable. As determined by simulations of Cr inclusions, volume of $0.3\%$ of the total cavity volume decreases the amplitude of the resonant peak by $55\%$, while higher concentration completely destroys the resonance, see Appendix. In comparison, Si nanostructures constitute $\sim6\%$ of the total cavity volume.

The nanostructured Si was embedded in a SiO$_2$ layer by atomic layer deposition (ALD) at a low growth rate of 1.19 \AA/cycle ensuring an air-gap-free cavity (Silayo ICP 330, Sentech Instruments GmbH). The Si nanostructures induce waviness in the upper layers, thus the deposited embedding layer was planarized by ion-beam etching (Oxford Ionfab 300, Oxford Instruments). The waviness was reduced to $A_{\text{w}}\approx15$~nm, which is significantly less than the operational wavelength and is not expected to deter the performance of the FP cavity, see Appendix. During the process, the physical length of the cavity was reduced to $910$~nm, less than the original design value of 950~nm.

Before the deposition of the top mirror, the surface of the SiO$_2$ cladding was pre-treated with Ar-ion plasma to create chemically active sites for better cross-link ~\cite{meyer1974}. Similarly to the bottom, the top mirror was optimized for high reflectivity at $\lambda = 1500$~nm. However, due to different exposure conditions, the thicknesses changed. In particular, the top DBR was constituted of 7 pairs of SiO$_2$ and TiO$_2$ layers with thicknesses of $t_{\text{2L}} = 250$~nm and $t_{\text{2H}} = 170$~nm, respectively, and the last layer of SiO$_2$ set to 100~nm. Because of the different thicknesses, the reflection band of the top mirror shifted, but high reflectivity ($R_{\text{m}}\approx98\%$, comparable to the initial design) in the spectral range of interest was maintained, see Appendix.

%%%%%%%%%%%%%%%%%%%%%%%%%%%%%%%%%%%%%%%%%%%%%%%%%%%%%%%%%%%%%%%%
\section{Results}\label{results}
\begin{figure}[t]
  \centering
  \includegraphics[width=1\linewidth]{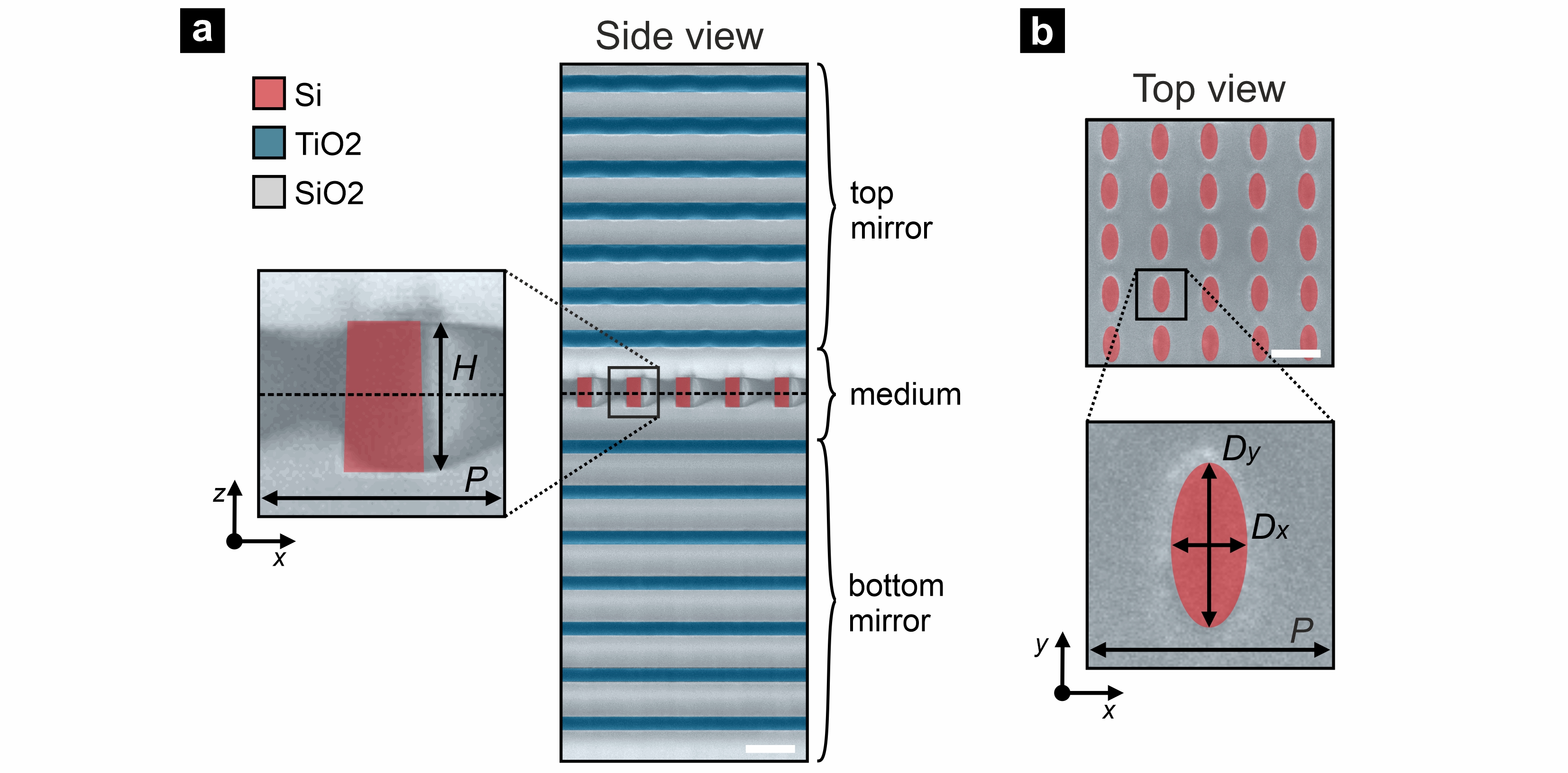}
  \caption{Exemplary pixel from the fabricated spectro-polarimetric system: 
  (a) side view, (b) top view. Colored SEM images of vertical and horizontal cross-sections of a pixel based on a FP cavity modulated by an array of Si nanostructures. Si is colored red, TiO$_2$ - blue, SiO$_2$ - grey. Si nanostructures are defined by their height ($H$) and diameter in $x$ and $y$ axis ($D_x$ and $D_y$, respectively), and are distributed in a square lattice with a period ($P$). Scale bars are equal to 500~nm.
  }\label{fig3}
\end{figure}

A focused ion beam (FIB) and a scanning electron microscope (SEM) were used to obtain vertical and horizontal cross-sections of the fabricated spectro-polarimeter, see Fig.~\ref{fig3}(a,b). The total thickness of the optical device is $7~\upmu$m. 

\begin{figure}[t]
\centering
    \includegraphics[width=1\linewidth]{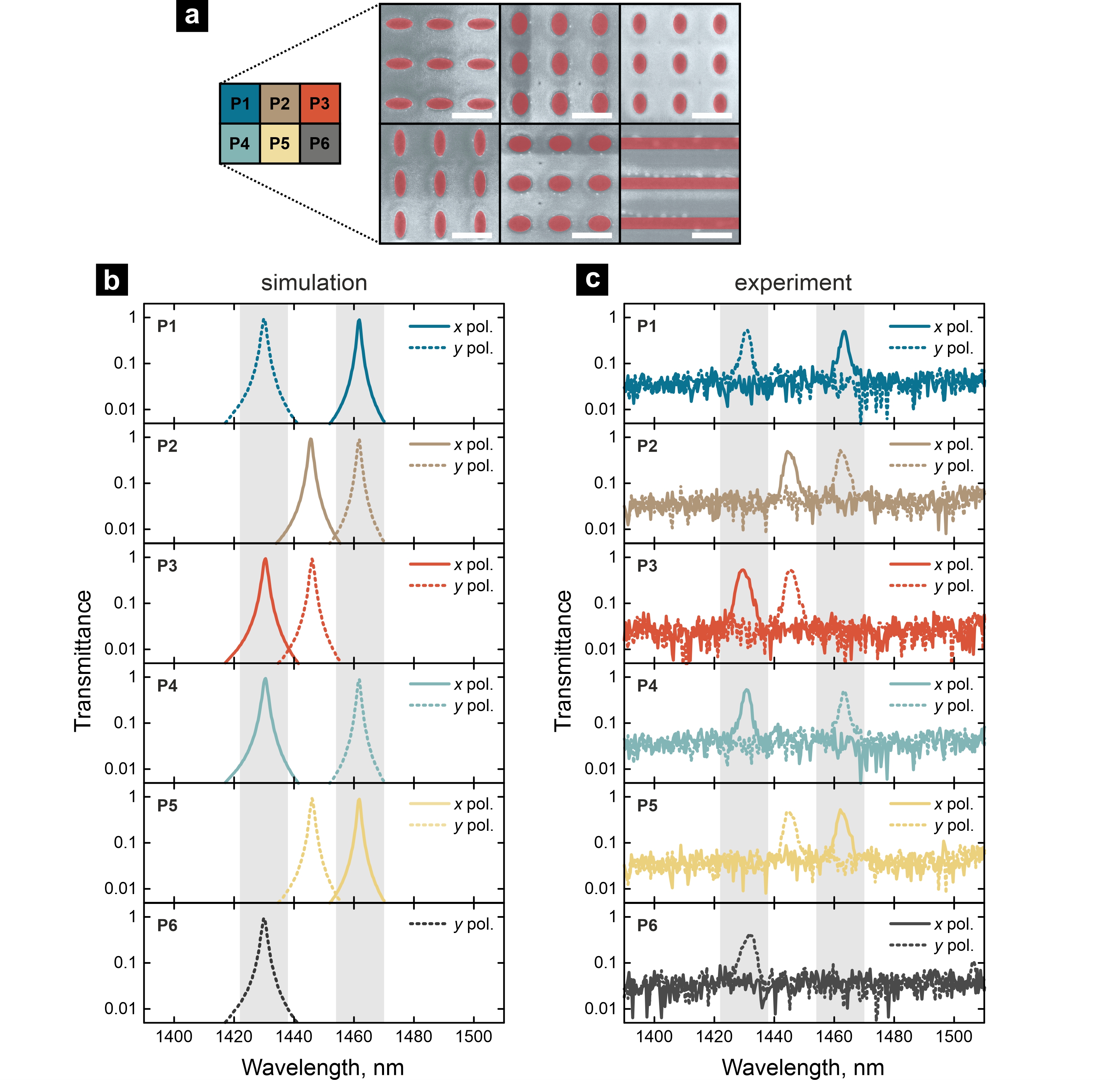}
  \caption{Pixels and their transmission spectra. (a) SEM images of horizontal cross-sections of all pixels ($\text{P1},\cdots,\text{P6}$), Si is colored red. Scale bar is equal to 500~nm. (b) Simulated transmission of 6 pixels, introduced in Fig. 2, with Lorentzian peaks centered at 1430~nm, 1446~nm, and 1462~nm and distributed in 3 spectral bands of $\Delta\lambda = 16$~nm. Peaks are (c) Measured transmission of the corresponding pixels. Spectra for $x$ polarized light is shown in solid lines, for $y$ polarized - dashed lines.}\label{fig4}
\end{figure}

Moreover, the horizontal cross-sections of all pixels ($\text{P1},\cdots,\text{P6}$) were imaged to evaluate the sizes of the nanostructures, see Fig.~\ref{fig4}(a). All of the nanostructures had a height of $H=300$~nm, but varied in both diameters, $D_x$ and $D_y$. The measured transverse dimensions of the nanostructures are provided in Tab.~\ref{tablesizes}. As can be seen from the standard deviation ($\sigma$) of the measured diameters , the proximity effect during electron beam exposure leads to some errors. However, it should be noted that as long as the error is homogeneously distributed in the whole measured area of $20\times20$ $~\upmu\text{m}^2$, it contributes mainly to the effective length of the cavity and not the amplitude or width of the transmission peaks. 

\begin{table}
\centering
\caption{Measured transverse dimensions ($D_{x,y} \pm \sigma$) of Si nanostructures in the set of 6 pixels and measured central wavelengths of their transmission peaks for $x$ and $y$ polarizations.}
\begin{tabular}{|l|l|l|l|l|}
\hline
 & $D_x$,~nm & $D_y$, nm & $\lambda_x$, nm & $\lambda_y$, nm  \\ \hline
Pixel 1 & $335\pm8$  & $145\pm6$  & 1463  & 1431  \\ \hline
Pixel 2 & $202\pm5$  & $282\pm5$  & 1445  & 1462  \\ \hline
Pixel 3 & $162\pm4$  & $247\pm4$ & 1430  & 1446  \\ \hline
Pixel 4 & $145\pm6$  & $335\pm8$  & 1431  & 1463  \\ \hline
Pixel 5 & $282\pm5$  & $202\pm5$  & 1462  & 1445  \\ \hline
Pixel 6 & -  & $134\pm5$  & -  &  1432 \\ \hline
\end{tabular}
\label{tablesizes}
\end{table}

The spectral measurements of the fabricated sample were carried out using a broadband halogen light source (SLS301, Thorlabs, Inc.). Its light was delivered to the sample via an optical system emulating the conditions of a plane-wave illumination. A linear polarizer was mounted on a rotational stage (PR50CC, Newport Corp.) in front of the sample to control the polarization state of the incident light. The sample was positioned and the position angle was calibrated using a 5-axis positioning system (Aerotech, Inc.). The transmitted light was collected via $20~\upmu$m aperture and a lens system to an optical spectrum analyzer with subnanometer spectral resolution (AQ6370B, Yokogawa). The measured transmittance of the six individual pixels, constituting elements of the super-pixel configuration, as shown in Fig.~\ref{fig4}(a), is presented and compared to simulation in Fig.~\ref{fig4}(b,c).

The measured peaks have a very good agreement with the simulation results regarding their central wavelengths. Accounting for the reduction of the cavity length compared to the initial design, the anticipated spectral positions of the peaks blue-shifted: $\lambda_1=1430$~nm, $\lambda_2=1446$~nm, $\lambda_3=1462$~nm. The central wavelengths of the measured peaks are given in Tab.~\ref{tablesizes}, the standard deviation is less than 1~nm. Moreover, the measured peaks reach a transmission of $46\%$, c, in the simulation the peak transmission is close to $92\%$. The width of the measured peaks are slightly broader, in average $\text{FWHM} = 3.6$~nm compared to $\text{FWHM} = 1.6$ in the case of the simulation. Such behavior is induced by the contamination of the cavity, as discussed in Sec.~\ref{fabrication}. Its impact on transmittance is discussed in Sec.~\ref{SNRdiscuss}. 

Now, in order to estimate the performance of the pixels, three spectral bands were selected, as described in Sec.~\ref{concept}. Since the measured spectral peaks are of a Lorentzian-shape, which has long tails, they were spaced $\Delta\lambda=16$~nm apart to minimize the cross coupling. Accordingly, the three spectral bands of the system were: 1422-1438~nm, 1438-1454~m, and 1454-1470~nm. As can be seen in Fig.~\ref{fig4}(c), the measurements have a high noise floor, which comes from the detector of the spectrometer and is not inherent to the fabricated structures. Thus, a prior analytical knowledge is used, and the intensity for a given band and polarization state is computed from the Lorentzian fit. This enables a significant reduction in noise. The elements of the reconstruction matrix are then obtained using the following integral:
\begin{equation}
\alpha=\int_{\lambda_{min}}^{\lambda_{max}} T_{pol}(\lambda)d\lambda,
\end{equation}
where $\lambda_{max}$ and $\lambda_{min}$ are the boundaries of the spectral band, and $T_{pol}(\lambda)$ is the transmittance of a pixel for a given polarization state. Computing these integrals for the measured data at all the spectral bands and polarization states results in the following matrix:
\begin{equation}
A= \left[\begin{array}{cccccc}
0.017&0.076&2.002&2.229&0.112&0.021\\
0.170&2.457&0.124&0.029&0.136&2.478\\
2.877&0.160&0.034&0.147&2.577&0.126\\
2.229&0.112&0.021&0.017&0.076&2.002\\
0.029&0.136&2.478&0.170&2.457&0.124\\
0&0&0&2.432&0.239&0.042\end{array}\right].
\end{equation}
This matrix has a condition number of $k(A)=8.43$, compared to ideal case $k(A)=8.0552$ (Eq.~(\ref{inversionmatrix}) $\alpha_m^n=1,\alpha_x^6=0$). The importance of this condition number is discussed in detail in the following section, see Sec.~\ref{discussion}.

\begin{figure}[t]
\centering
\includegraphics[width=1\linewidth]{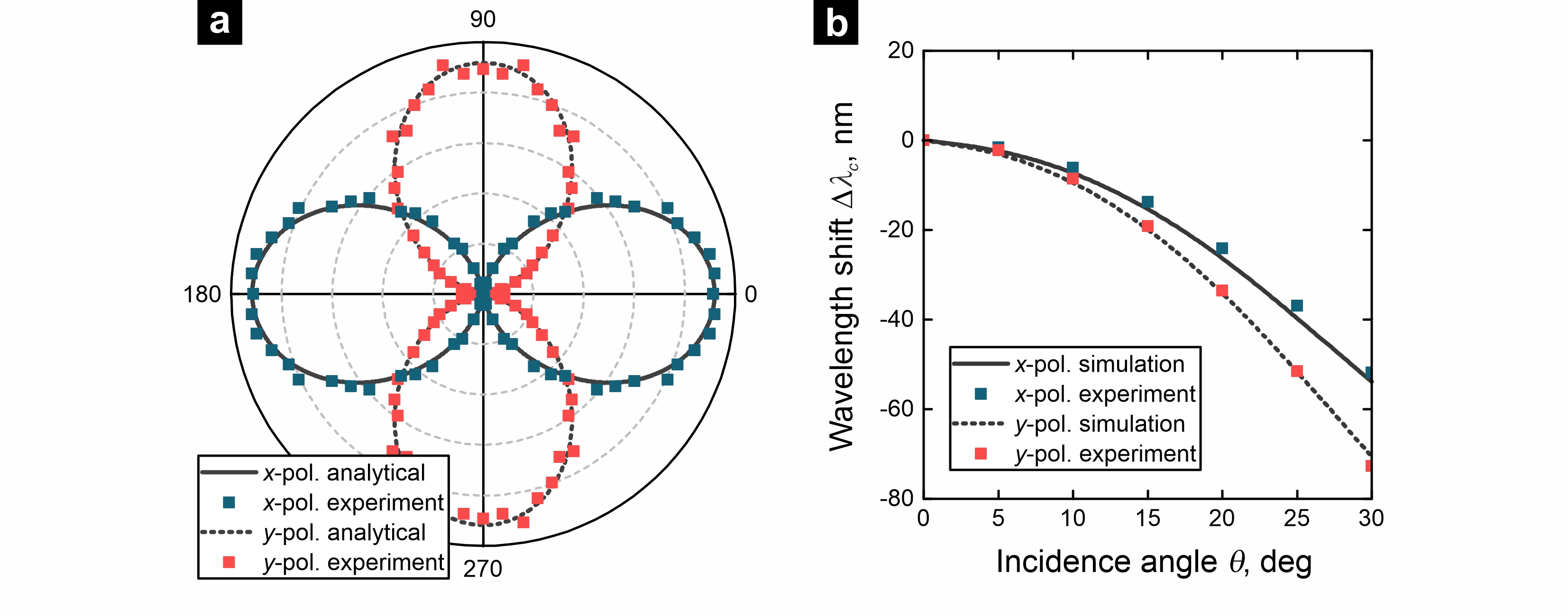}
\caption{Dependence on polarization angle $\phi$ and incident angle $\theta$. (a) Normalized measured intensity at $\lambda_x$ and $\lambda_y$ as a function of polarization angle $\phi$ in case of arbitrarily selected pixel with nanostructure inclusion. The intensity follows analytically predicted $\cos^2{\phi}$ and $\sin^2{\phi}$ functions, respectively. (b) Blue-shift of the central wavelength of the peak $\lambda_{\text{c}}$ for $x$- and $y$-polarization in case of an empty cavity. Simulations are shown in solid and dashed line, while the experimental points are depicted by blue and red squares, respectively.
} \label{fig5}
\end{figure}

%This increase of the condition number leads to a very slight decrease of the signal to noise ration $\mathrm{SNR_{real}}=\mathrm{SNR_{ideal}}-\{0.176\}dB$ with respect to the ideal case.
In addition, we measured the systems response to a change of the azimuthal angle $\phi$ and polar angle $\theta$ of the incident light, see Fig.~\ref{fig5}(a,b). First, as two different peaks centered at $\lambda_x$ and $\lambda_y$ are produced for two linear orthogonal polarization states, $x$ and $y$, respectively, we measured the intensity at $\lambda_x$ and $\lambda_y$ as a function of azimuthal angle $\phi$. As can be seen from Fig.~\ref{fig5}(a), the intensity follows the analytically predicted $\cos^2{\phi}$ and $\sin^2{\phi}$ functions, respectively. Second, multi-layer systems are known for high angle dependence. In Fig.~\ref{fig5}(b) we observe a blue-shift of the central peak of the transmission of an empty cavity with the increase of the polar angle $\theta$ in $x$-axis with respect to the surface normal for $x$ and $y$ polarization in both experimental and numerical data. Quantitatively comparable dependence was numerically observed for all of the pixels. Accordingly, with the increase of numerical aperture (NA) of the imaging system in front of the spectro-polarimeter, the peak is expected to broaden ~\cite{atherton1981}. Further discussion on impact of incidence angle and subsequent assumptions are given in Section 4.5.

%%%%%%%%%%%%%%%%%%%%%%%%%%%%%%%%%%%%%%%%%%%%%%%%%%%%%%%
\section{Discussion} \label{discussion}
%In this section the boundaries on the performance of the presented spectro-polarimetric sensor will be discussed and using these limitation a optimal system and its expected performance will be derived.

In this section, we will discuss the performance of the presented spectro-polarimetric sensor, and extend the concept to an optimal system.

\subsection{Signal-to-noise considerations}
\label{SNRdiscuss}
In order to assess the performance of our spectro-polarimetric system, it is important to take a look at the noise propagation through the system. This allows an estimation of the SNR of the Stokes parameters as function of the pixel SNR of the sensor. For this analysis the condition number $k(A)$ is used, since it is a measure of the sensitivity to variations for a standard system of equations $A\mathbf{x}=\mathbf{b}$. The condition number $k(A)$ is defined by:
 \begin{equation}
 \frac{||\delta \mathbf{x}||}{||\mathbf{x}||}=k(A)\frac{||\delta\mathbf{b}||}{||\mathbf{b}||},
 \label{cond}
 \end{equation}
where $\delta \mathbf{x}$ and $\delta \mathbf{b}$ are small variations on the corresponding vectors $\mathbf{x}$ and $\mathbf{b}$. A common way to compute the condition number of matrix $A$, is to take the ratio of the largest singular value and the smallest singular value of the matrix $A$. This method is used to compute the condition number in the paper. 
As can be seen from Eq.~(\ref{cond}), the condition number $k(A)$ expresses the proportionality between any variations of the known vector $\mathbf{b}$ and the unknown vector $\mathbf{x}$. Thus, in the presented system, it relates the noise of the measured intensities (shot noise, read-out noise, etc.) to the noise of the reconstructed spectro-polarimetric intensities $I_{pol}^{\lambda_n}$. Using the relation between the polarized intensities and Stokes parameters (Eqs.~(\ref{S0}-\ref{S2})) the expected value ($\mathrm(E)(x)$) and standard deviation ($\sigma(x)$) of the reconstructed Stokes parameters can be derived as reported here (seen Appendix for derivation):
\begin{align}
\label{S0est}
\mathrm{E}(\tilde{S_0})&=S_0\\
\label{S1est}
\mathrm{E}(\tilde{S_1})&=S_1\\
\label{S2est}
\mathrm{E}(\tilde{S_2})&=S_2\\
\sigma(\tilde{S_0})&=\left(\frac{k(A)}{\mathrm{SNR}}\right)\\
\sigma(\tilde{S_1})&=\sqrt{2}\left(\frac{k(A)}{\mathrm{SNR}}\right)\\
\sigma(\tilde{S_2})&=\sqrt{2}\left(\frac{k(A)}{\mathrm{SNR}}\right).
\end{align}
From Eqs.~(\ref{S0est}-\ref{S2est}) it can be seen that the reconstruction method is bias-free, since the expected values are equal to the true values. Note that the standard deviation of $\tilde{S_0}$ is smaller than of other Stokes parameters due to the fact that $\tilde{S_0}$ is computed/measured twice according to Eq.~(\ref{S0}). Since the standard deviation is inversely proportional to the SNR, the SNR after reconstruction will be reduced by a factor $k(A)$. For the system of equations shown in Eq.~(\ref{inversionmatrix}), using $\alpha^{6}_{x}=0$ and setting all others to $\alpha^{\lambda}_{p}=1$, one obtains $k(A)=8.0552$. Thus the noise in the reconstructed signal will be increased by at most this factor. In comparison, the polarimeter in \cite{Wei2017} uses a metasurface as polarization sensitive lens. The different polarization states are focused on different parts of the sensor, thus the metasurface is spatially separated from the sensor by the focal length. This system has a condition number of $k(A)=3.6581$. Another example is the spectro-polarimeter in \cite{Chen2016}, which achieves a condition number of $k(A)=2.082$. However, an external spectrometer is needed in their setup. Despite the higher condition number, the nanostructure-modulated spectro-polarimeter presented in this paper can be directly integrated on top of a sensor and it does not require an external spectrometer, resulting in a very compact system.

%\begin{itemize}
%   \item explain shifted peaks --> lower cavity
%   \item explain lower peaks --> chromium resist contamination
%	\item interpretation of change in SNR and effect of lower peaks on SNR --> change in $k(A)$ and reduction of SNR due to low T
%	\item interpretation on maximum bandwidth (FB limit,separation limit) of system/ maximum number bands
%	\item change expected change SNR due scaling size problem
%	\item angle sensitivity
%\end{itemize}
%As can be seen in Fig.\ref{fig2} the  measured transmission peaks are a lower than the simulated ones this is likely caused by some left over chromium that is present on top of the resonators. This leftover material can be seen in Fig.\ref{contamination} \textbf{(need clear picture of stuff on top of resonators)}, this small amount of lossy material has a high impact on the transmission due to its placement in a high quality cavity. The loss inside the cavity will have an impact on the SNR of the Stokes parameters. 

Another factor that must be taken into account in the evaluation of the system SNR is the reduced transmittance and the presence of shot noise. The latter is dependent on the total number of photons reaching the detector $N_{\text{phot}}$. In particular, the SNR of a shot noise limited pixel is given by: $\mathrm{SNR_{shot}}=\sqrt{N_{\text{phot}}}$. Thus, reducing the number of photons that reaches the detector by a factor of 2 (50\% transmission), the SNR is reduced by a factor of $\sqrt{0.5}$. Therefore, a system based on the measured structures would have a SNR reduced by a -1.681~dB, compared to an ideal system. However, the effect of this low transmission can be reduced by doubling the measurement duration (the integration time of the detector).

\subsection{Bandwidth considerations}
Other than the SNR of the system, the achievable total bandwidth is also of importance for the applications of interest. The bandwidth of this presented spectro-polarimetric system depends on the reflection band of the mirrors and the maximum achievable separation of the transmission peaks. The relative bandwidth ($\frac{\Delta f}{f_0}$) of a DBR-mirror using quarter-wavelength sections is given by\cite{Orfandidis2008}:
\begin{equation}
\frac{\Delta f}{f_0}=\frac{4}{\pi}\arcsin(\rho),
\end{equation}
with $\rho$ being the Fresnel reflection coefficient. From this equation, it is clear that the bandwidth of the mirror depends only on the difference betwen the refractive indices of the two materials used in the construction of DBRs. In our case, the mirrors limit the bandwidth to $\sim{400}$~nm, see Appendix, but using more advanced mirror designs\cite{Zhong2015}, it is possible to obtain a wider reflection band. Therefore, bandwidth limitations due to the mirrors can be mitigated. In practise, the total bandwidth is limited by the maximum achievable separation of two linear orthogonal polarization peaks. As illustrated in Appendix, the maximum separation of the two peaks is equal to 127~nm. It can be achieved with a subdiffractive grating of 200~nm width. This separation between the peaks ultimately determines the maximum achievable total bandwidth of the system, since one of the transmission peaks has to be outside of the reconstruction bands. This results in the one row of the matrix having only one peak, which is the necessary condition for the spectro-polarimetric retrieval. 
The maximum number of bands can now be computed using:
\begin{equation}
N=\left\lfloor\frac{\mathrm{BW}}{\Delta\lambda}\right\rfloor-1,
\end{equation}
where BW is the total bandwidth of the system and $\Delta\lambda$ is the width of a single spectral band.

\subsection{Spectral resolution}
\begin{figure}[t]
\centering
\includegraphics[width=1\linewidth]{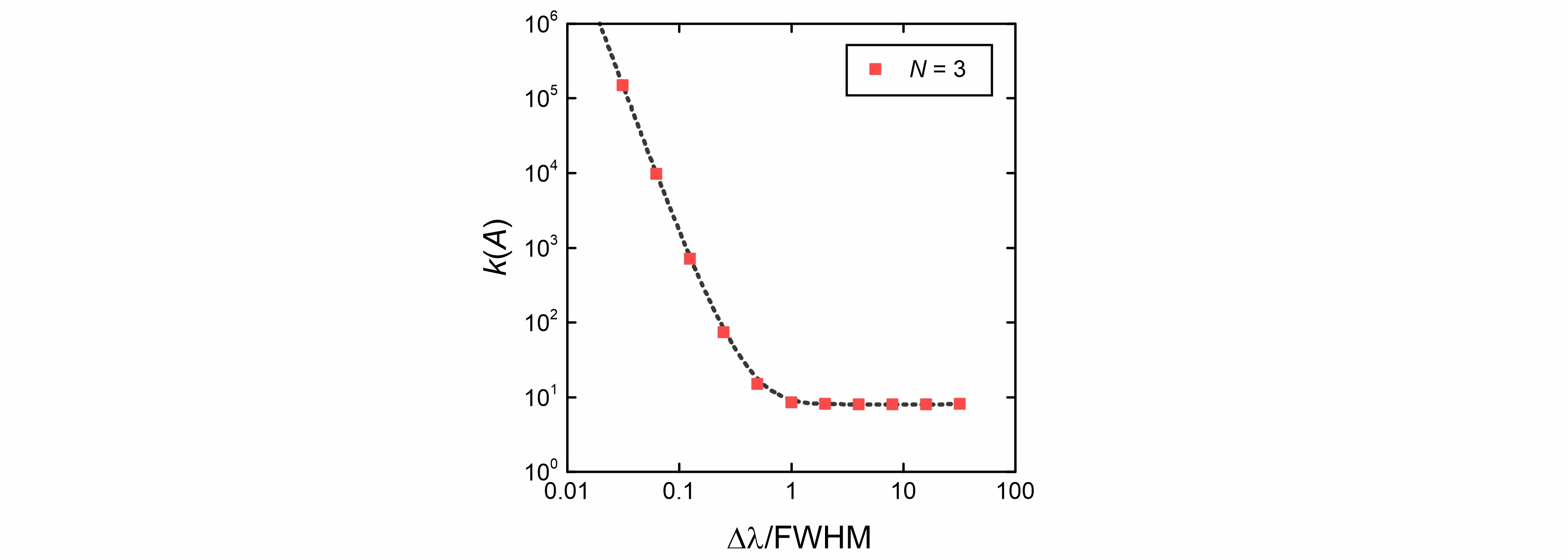}
\caption{Condition number $k(A)$ vs spectral resolution. Condition number $k(A)$ of a reconstruction matrix with 3 bands is shown in cases of different ratios between the single bandwidth $\Delta\lambda$ and the transmission peak width FWHM, assuming a Lorentzian-shape peaks. }\label{fig6}
\end{figure}

In the presented system, the spectral positions and the width of the spectral bands have to be chosen carefully to obtain the optimal performance. By decreasing the width of a single band, the number of bands can be increased, and thus its spectral resolution. However, as can be seen from Fig.~\ref{fig6}, reducing the width of the spectral band $\Delta\lambda$ below the FWHM of the transmission peaks greatly increases the condition number and in turn the SNR of the system. This increase of the condition number $k(A)$ can be attributed to the spectral overlap of the neighboring peaks. This translates to smaller differences between the rows of the reconstruction matrix, making the system of equations more linear dependent.
It is recommended to use $\Delta\lambda=\mathrm{FWHM}$, limiting the effective spectral resolution to FWHM.
% maybe say something about realistically achievable FWHM.

\subsection{Handling a large number of bands}
\begin{figure}[t]
\centering
\includegraphics[width=1\linewidth]{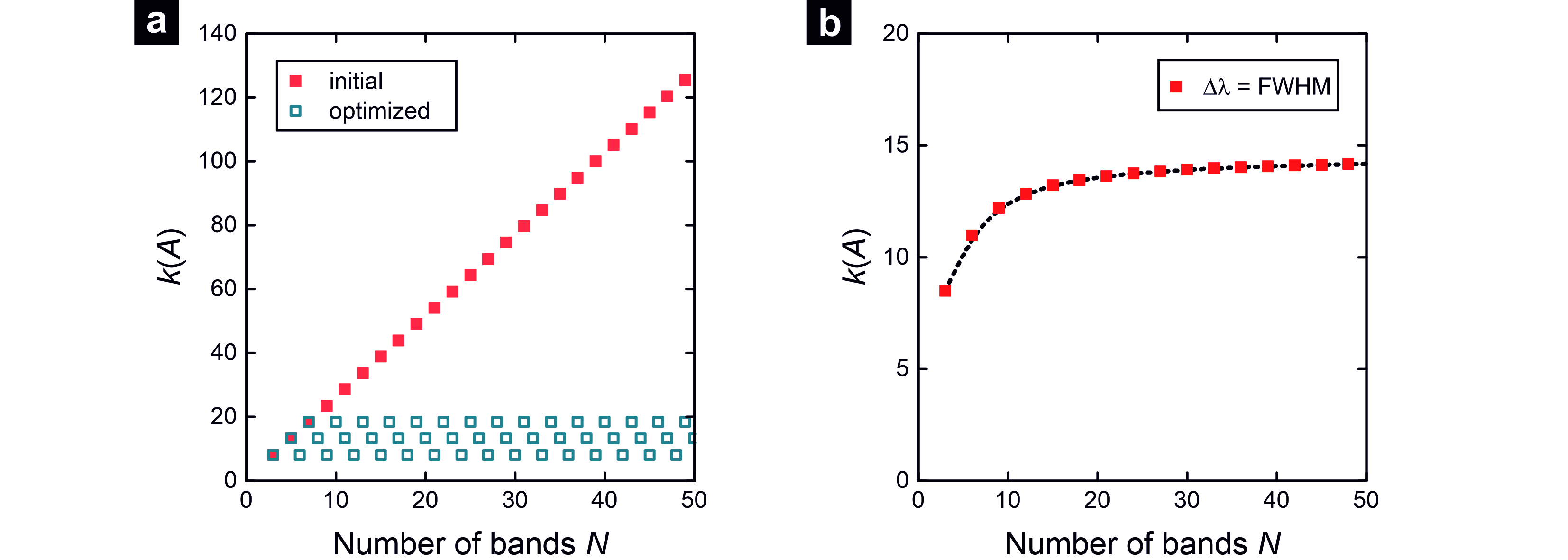}
\caption{Condition number $k(A)$ vs number of spectral bands $N$. (a) $k(A)$ of a reconstruction matrix for N bands, assuming all elements of a unit magnitude at the peaks and zeros elsewhere. Red squares represent a direct inversion using a certain number of pixels and only a single pixel with a grating. Empty blue squares represent an inversion after segmenting the matrix in submatrixes. (b) $k(A)$ of a reconstruction matrix with $N$ bands for $\Delta\lambda$ equal to the FWHM, assuming a Lorentzian-shape peaks, as measured.}\label{fig7}
\end{figure}

Another effect that must be taken into account in selecting the number of bands, is the rank of the reconstruction matrix. The matrix without the grating will have a rank deficiency of 1 for an odd number of bands while the rank deficiency for an even number of bands is 2. Thus, for an even number of bands replacing a pixel with a grating cannot make the matrix invertable. Accordingly, to retrieve the polarization state for all bands, the number of bands must be segmented in sets of an odd number of bands. Since for a system of equations with an odd number of bands the condition number $k(A)$ changes linearly with the number of bands, see Fig~\ref{fig7}(a), the reconstruction matrix should be constructed from as small as possible submatrixes. 
Any number of bands greater than 4 can be written as a sum of 3s, 5s and 7s, so is possible to segment the reconstruction matrix in submatrixes with these numbers of bands. In that case, the worst case condition number of the total matrix would be that of the 7 band inversion, $k(A)=18.36$ (under the assumption of the matrix with all unit $\alpha$'s except for the one that is 0). The behavior of such an optimized matrix is shown in Fig.~\ref{fig7}(a). Still, it is recommended to use a system with a number of bands that is divisible by 3, since such a system will have a SNR ratio that is more than double of that of the presented worst case (7 bands) scenario.

In Fig.~\ref{fig7}(b), the condition number $k(A)$ is plotted for segmented systems (block of 3 spectral bands) when Lorentzian peaks are used to compute the transmission values in the matrix, thus also taking into account the spectral cross-coupling between all pixels. Here, a single bandwidth was chosen to be equal to the FWHM of the transmission peaks, since it is the limit for a low condition number $k(A)$, as shown in Fig.~\ref{fig6}. Because the tails of the Lorentzian functions leak into the bands of adjacent submatrixes, they become partially dependent. The condition number of the matrix increases, but the segmenting still significantly reduced the condition number $k(A)$ compared to the unsegmented initial matrix in Fig.~\ref{fig7}(a). The plotted data of Fig.~\ref{fig7}(b) is fitted by a curve:
\begin{equation}
k(A)=-15.69N^{-0.8}+14.90~,
\label{fit}
\end{equation}
resulting in fit with a coefficient of determination of $R^2=0.9975$. Based on this fit, the condition number $k(A)$ will converge to a value of 14.90 for a large number of bands.

\subsection{Optimal design}
Using the design limitations discussed previously, as an example of ideal design, we can consider a spectro-polarimetric sensor with 126 bands (42 x 3), the bandwidth of 127~nm and a spectral resolution of 1~nm, centered around the $\lambda=1500$~nm. Then the system that can retrieve the first three Stokes parameters ($S_1$, $S_2$, $S_3$) would consist of 504 pixels and would have a condition number of 14.57. This results in a reconstruction of the $S_0$ with a SNR that is -11.63 dB below the shot noise limited SNR. The $S_1$  and $S_2$ would be reconstructed with a SNR that is -13.14 dB below the noise shot limited SNR.

When implementing this design in an imaging device the chief ray angle (CRA) across the detector surface has to be taken into consideration. For a conventional optical system this angle will increase radially from the center of the detector. This increasing incident angle leads to a blue-shift of the transmission peaks. In principle, this spectral shift can be compensated by scaling the effective refractive index of the cavity, e.g. scaling the size of the high-index inclusions \cite{Frey2015}. Alternatively, a telecentric optical system can be used. In such a system the CRA is constant over the entire detector and thus no blue-shift occurs. Furthermore, a system with a low NA is recommend since a high NA will increase the transmission peak width \cite{atherton1981}, thus reducing the spectral resolution of the system. 

The presence of a focused beam will influence the polarization reconstruction. When a fully polarized beam is focused by an optical system some of the light will be cross polarized. 
For example, the far-field of a focused beam arising from a microscope objective, illuminated by a fully x-polarized incident beam, is given by \cite{novotny2006}:
\begin{equation}
\mathbf{E_\infty}=E_{inc}\frac{1}{2}\left[
\begin{array}{c}
\left(1+\cos\theta\right)-\left(1-\cos\theta\right)\cos\left(2\phi\right)\\
-\left(1-\cos\theta\right)\sin\left(2\phi\right)\\
-2\cos\phi\sin\theta
\end{array}
\right]
\sqrt{\frac{n_1}{n_2}}\left(\cos\theta\right)^{\frac{1}{2}},
\label{depolarization}
\end{equation}
where $\mathbf{E_\infty}$ is the electric far-field of the focused beam in Cartesian coordinates, $E_{inc}$ the amplitude of the incident x polarized beam, $n_1$ and $n_2$ the refractive indices of the media before and after the lens and $\phi$, $\theta$ the azimuthal and polar angles of the far-field. As can be seen from this equation, the cross polarized component increases with the polar angle.This change in polarization can be interpreted as a depolarization effect that occurs before the planar spectro-polarimeter. Thus, the reconstructed polarization state will have a lower degree of polarization compared to the incident beam. Based on Eq.~(\ref{depolarization}), a maximum coupling from the x to the y component of 0.7\% is expected, at a polar angle of $10^\circ$. From a more advanced analysis that takes into account the full coupling within the structure, a higher value of 2\% is obtained. This detailed theoretical analysis is not reported here for sake of brevity and will be the topic of another extended paper. From these considerations, it is clear that in order to minimize this effect, a telecentric and low NA optical system is recommended.

Finally, we present a few considerations on the impact of the proposed concept on the spatial resolution achievable in a system with a given number of pixels. As discussed in the previous example, in order to retrieve the first three Stokes parameters over 3 spectral bands, 12 pixels have to be used. Thus, for comparisons sake, we can say that, from the spatial resolution point of view, 4 pixels are required per spectral band. In typical imaging multi-spectral systems, only one pixel per spectral band is used. Therefore, for the same number of sensor pixels, the spatial resolution of the proposed system is a quarter of that of a conventional multi-spectral camera, without any polarimetric functionality. On the contrary, in a typical polarimeter, 4 pixels are used to determine the polarization state. As a consequence, we can conclude that the effect on the spatial resolution of the presented system is comparable
to that of a typical spectro-polarimeter
obtained by combining "conventional" spectral and polarimetric
components. 

%COMMENT: In RGB spetropolarimeter some actually use Bayer pattern + 4 polarizers per color, in total 16 (https://thinklucid.com/phoenix-polarization-color-camera-featuring-sony-imx250myr-sensor-at-cvpr-2018/), we can actually mention that; but mostly it makes sense to compare to what Stefan had: [greyscale] with 4, thus 12 for 3 bands (https://thinklucid.com/tech-briefs/polarization-explained-sony-polarized-sensor/). 

%%%%%%%%%%%%%%%%%%%%%%%%%%%%%%%%%%%%%%%%%%%%%%%%%%%%%%%%%%%%
\section{Conclusion}
In this work, we have shown the feasibility of a planar spectro-polarimeter. As a proof of concept, we have experimentally demonstrated a set of 6 pixels with transmission peaks of 50\% and $\text{FWHM}=3.6$~nm. The peaks were separated in three spectral bands of $\Delta\lambda=16$~nm and sorted by their polarization state. Using the measured data in the reconstruction matrix, we obtained a condition number of $k(A)=8.43$, which is extremely close to the theoretical limit of $k(A)=8.06$. Such experimental results permit the reconstruction of the first 3 Stokes parameters up to a level of -10.76 dB below the shot noise limited SNR.

In addition, the limits of the proposed spectro-polarimetric design were analyzed with respect to the highest number of bands possible and the highest obtainable spectral resolution. The total system bandwidth of the current architecture is limited to 127~nm. The maximum condition number, limiting the SNR of the reconstruction, is estimated to be in the order of 14.57, given that the transmission peaks are separated by their FWHM. Having a spectral resolution of 1~nm, such system could have a bandwidth of 127~nm separated in 126 bands. The SNR of the Stokes parameters would be -13.14 dB below the shot noise limited SNR. 

In perspective, the spectral resolution of the system is limited only by the reflectivity of the mirrors. Thus, subnanometer resolution is possible. 
Moreover, the design could be scaled to other spectral ranges with respective changes in material selection, e.g. using TiO$_2$ instead of Si in visible spectral range to minimize intrinsic losses. Using several different cavities at once, this would allow a very broadband sensor. 
Also, the system would benefit from a more anisotropic inclusion inside the cavity in place of the grating in order to increase the bandwidth. Finally, the system could be extended with additional layers of retarders to enable the measurements of circularly polarized light, enabling the retrieval of the full Stokes vector.

%In addition, the inclusion of the nanostructures enables the fabrication of the spectro-polarimetric system with a uniform height in contrast to a step-profile as in conventional FP filters~\cite{Wang2018}.

\newpage
\section*{Appendix}

%In this version of the paper we attached Supplementary material. In case of Optica a separate document would be created, in case of Optics Express the material would be added as appendixes. At the moment, only figures are added for better understanding of the claims made in the main body of the paper. 

\subsection*{Modulation via transverse dimensions}
\begin{figure}[h]
\centering
\includegraphics[width=1\linewidth]{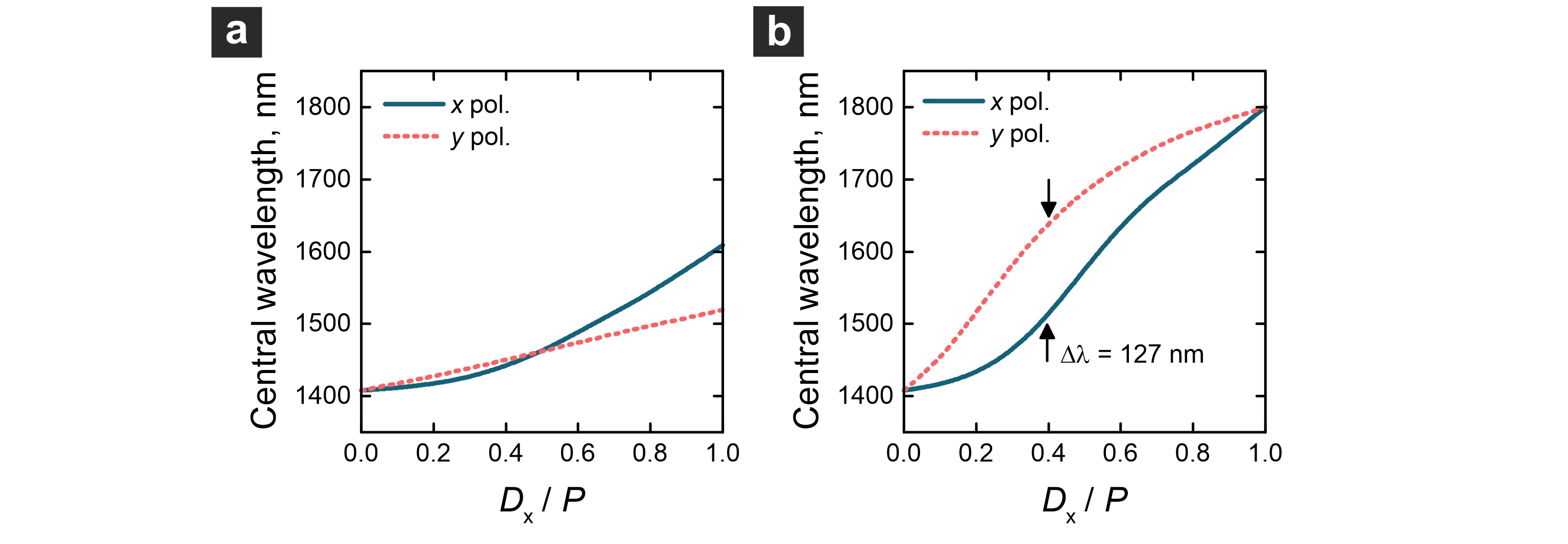}
\caption{Spectral positions of transmission peaks for $x$ and $y$ polarization as a function of width $D_x$ of the inclusions. (a) A case of an array constituted of elliptical nanostructures with $D_y=250$~nm, while $D_x$ was varied from 0~nm to 500~nm. (b) A case of a linear grating with its width varied from 0~nm to 500~nm (equal to period $P$). The maximum separation of the peaks is 127~nm.}\label{figS1}
\end{figure}

The central wavelength $\lambda_\text{c}$ changes with respect to the size of the inclusion. We show the control of $\lambda_\text{c}$ for $x$- and $y$-polarization by changing the diameter in $x$-axis for two different cases: a polarization-sensitive inclusion and a grating, see Fig.~\ref{figS1}(a) and Fig.~\ref{figS1}(b), respectively.

\subsection*{Reflectivity and bandwidth of DBRs}
\begin{figure}[h]
\centering
\includegraphics[width=1\linewidth]{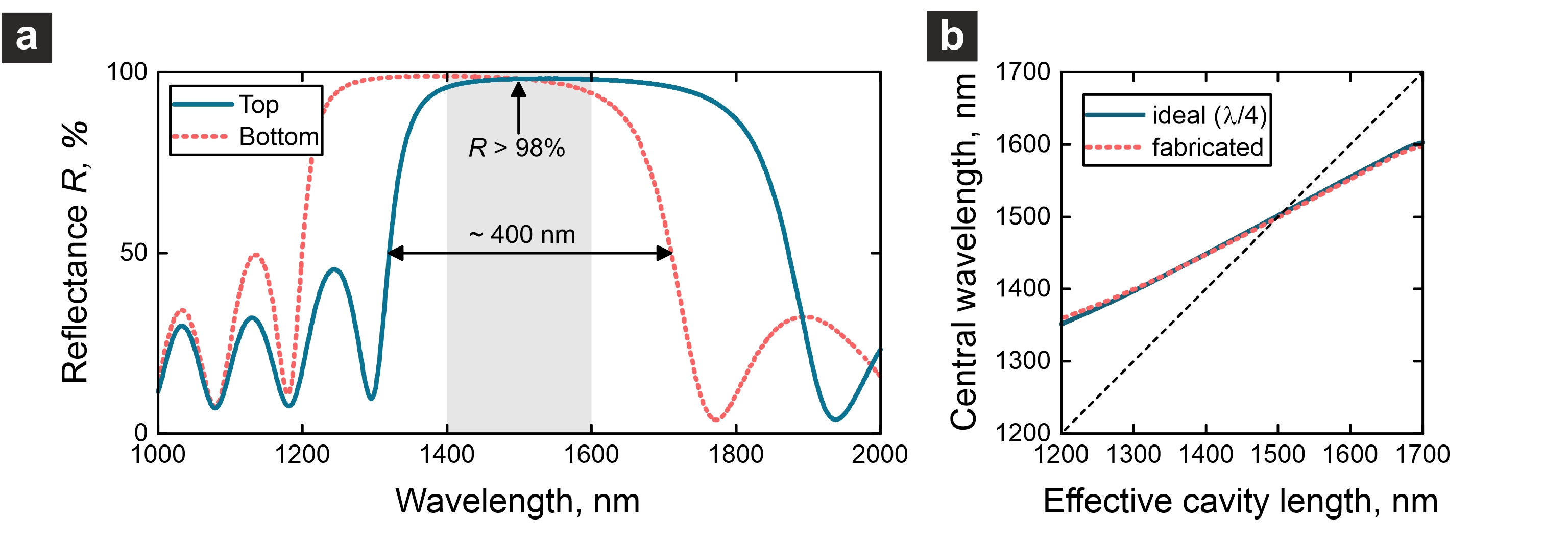}
\caption{Reflectivity of the fabricated mirrors and their impact on the FP resonance. (a) Measured reflection spectra of the bottom and the top DBRs. Bandwidth and reflectance in the spectral range of interest is highlighted. Due to the mismatch of mirrors, the bandwidth is smaller than of an individual mirror. (b) Simulated central positions of the FP resonance depending on the effective length of the cavity, when different mirrors are used: ideal ($\lambda/4$, for 1500~nm), and fabricated. For comparison, a dashed line represents a case for which mirrors and effective length of the cavity are $\lambda/4$ and $\lambda/2$, respectively.}\label{figS2}
\end{figure}

The fabricated DBR mirrors have slightly different properties due to different deposition conditions, as discussed in Sec.~\ref{fabrication}. Regardless of that, both mirrors have a high-reflectivity at $\lambda = 1500$~nm, see Fig.~\ref{figS2}(a). Despite the fact that mirrors are different from each other and the ideal $\lambda/4$ case, their impact on the central wavelength $\lambda_\text{c}$ is negligible, see Fig.~\ref{figS2}(b).

\newpage
\subsection*{Evaluation of contamination and waviness}
\begin{figure}[ht]
\centering
\includegraphics[width=1\linewidth]{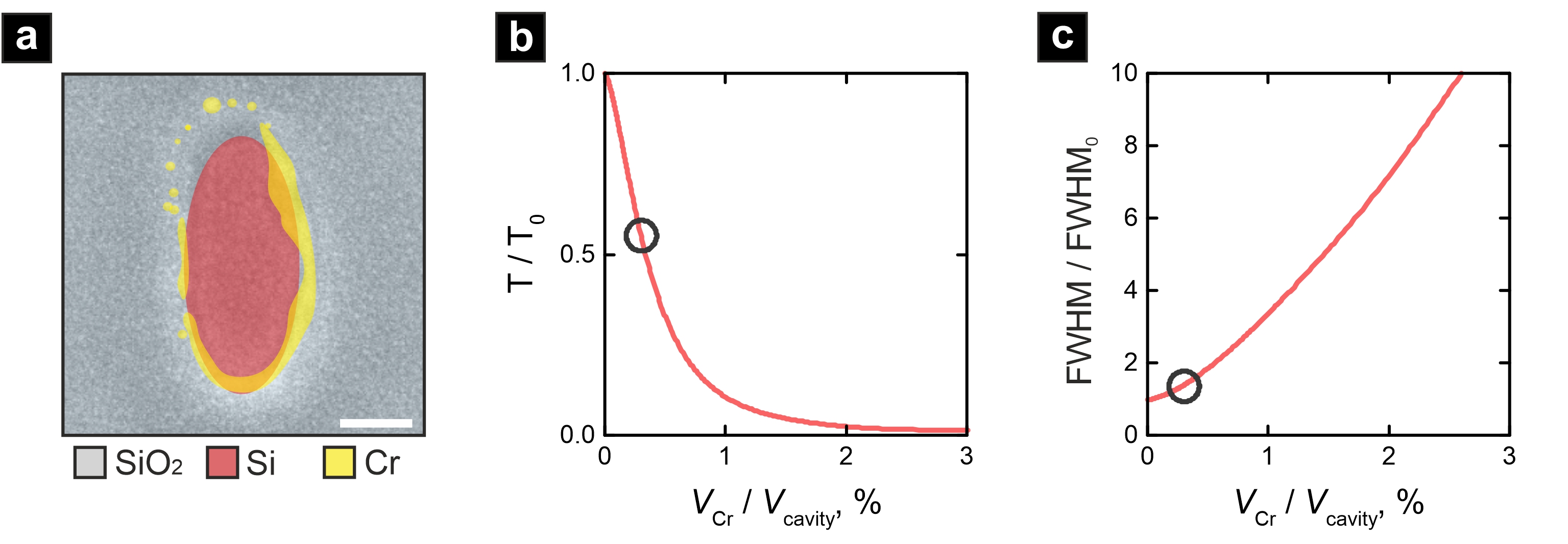}
\caption{Impact of cavity contamination with Cr.
(a) Colored SEM image of horizontal cross-section of a single elliptical nanostructure inside the cavity. Si$_2$ is red, SiO$_2$ is grey, and Cr is yellow. (b) Simulated relative peak transmittance (intensity) as a function of volume of Cr in the cavity. Cr is considered elliptical for simplicity of the model. (c) Simulated relative peak width (FWHM) as a function of volume of Cr in the cavity. Circle highlights anticipated value.}\label{figS3}
\end{figure}

\begin{figure}[ht]
\centering
\includegraphics[width=1\linewidth]{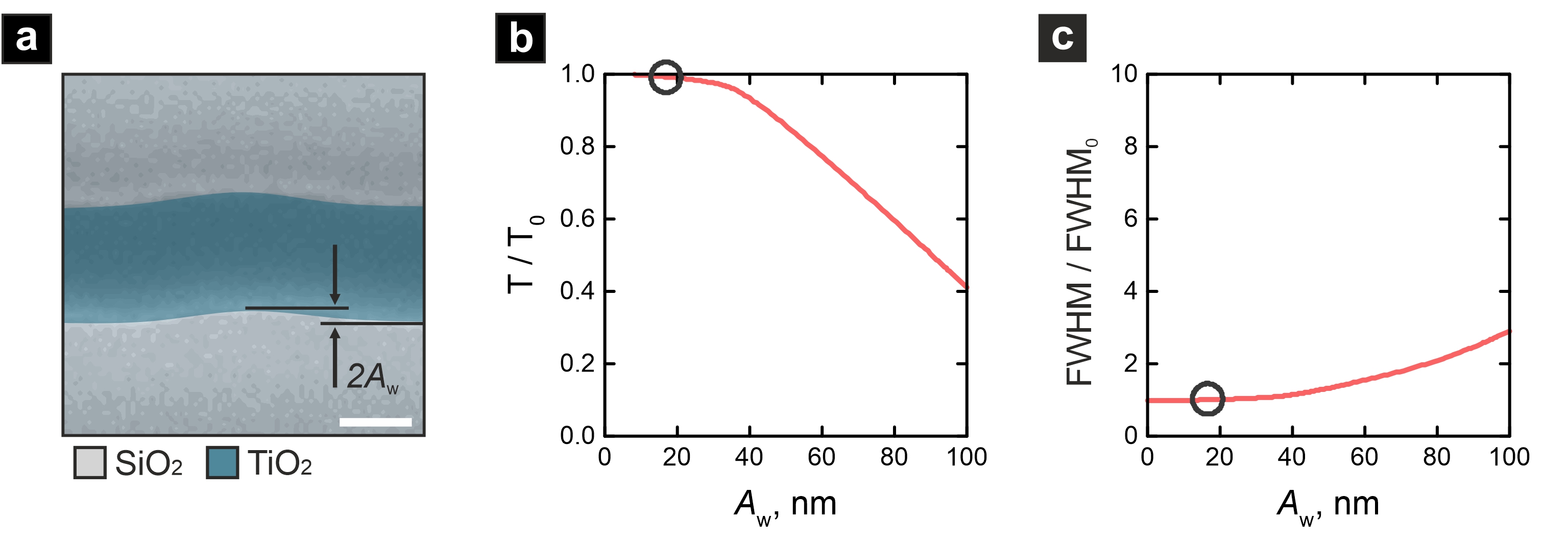}
\caption{Impact of top mirror waviness. (a) Colored SEM image of vertical cross-section of a single high-index DBR layer. TiO$_2$ is blue, SiO$_2$ is grey. The uneven surface can be desribed as sinusoidal surface with $A_\text{w} = 15$~nm. Scale bars is equal to 100 nm. (b) Simulated relative peak transmittance (intensity) as a function of waviness amplitude. (c) Simulated relative peak width (FWHM) as a function of waviness amplitude. Circle highlights anticipated value based on experimentally obtained $A_\text{w} = 15$~nm.}\label{figS4}
\end{figure}

The fabrication endured several challenges, as discussed in Sec.~\ref{fabrication}. Here we show the numerical simulation results of the impact of Cr contamination of the cavity, see Fig.~\ref{figS3}, and the waviness of the top mirror, see Fig.~\ref{figS4}. 
As illustrated in Fig.~\ref{figS3}(a), some amount of Cr was left in the cavity, approx. 0.3~\% of the total cavity volume.  Cr has a high absorption in the visible spectral range, thus the transmittance of the cavity rapidly decreases with increase of the Cr volume in the cavity, see Fig.~\ref{figS3}(b), while the FWHM increases, see Fig.~\ref{figS3}(c). 
The nanostructures induce waviness of the layers on top. Even after the planarization, the waviness remain of an amplitude $A_\text{w}=15$~nm, see Fig.~\ref{figS4}(a). In general, our simulations show that waviness may decrease the transmittance and increase the FWHM, as shown in Fig.~\ref{figS4}(b) and Fig.~\ref{figS4}(c), respectively.

\subsection*{Derivation of signal-to-noise ratio of Stokes parameters}
In order to derive the SNR of the Stokes parameters, the noise on the measured intensities first has to be defined. A convenient choice is to assume unit magnitude intensity on the pixels with some additive zero mean Gaussian noise. Using this normalized intensity, the SNR is inversely proportional to the normalised standard deviation ($\sigma$) of the Gaussian distribution ($\mathcal{N}(\mu,\sigma^2)$). This normalization aids to simplify the derivation. Thus, the intensity of the pixel is given by:
\begin{equation}
    I=I_s+\delta I=1+\mathcal{N}\left(0,\frac{1}{\mathrm{SNR}^2}\right),
\end{equation}
where $I_s$ is the signal intensity with a value of 1 and $\delta I$ is the noise intensity equal to the Gaussian noise. If it is now assumed that all pixels receive equal intensity (uniform assumption on polarization state and spectral signal), the expected length of the measured intensity vector $\mathbf{I}$ ($E(||\mathbf{x}||)$ with $||\mathbf{x}||$ being the 2-norm of the vector) can be computed, it is needed later in the derivation. The intensity vector is once again split in a signal part $\mathbf{I_s}$ and a noise part $\mathbf\delta{I}$, and can be written as:
\begin{equation}
    \mathbf{I}=\mathbf{I_s}+\delta\mathbf{I}=\mathbf{1}+\mathcal{N}\left(\mathbf{0},\frac{1}{\mathrm{\mathbf{SNR}}^2}\right),
\end{equation}
where $\mathbf{1}$, $\mathbf{0}$ and $\mathrm{\mathbf{SNR}}^2$ are vectors with ones, zeros or the  $\frac{1}{\mathrm{SNR}^2}$ as elements. In order to compute the expected length of the measured intensity vector, the expected length of the signal part and noise part are computed separately. 
For the signal intensity vector the length is simply given by:
\begin{equation}
    ||\mathbf{I_s}||=\sqrt{N},
\end{equation}
where $N$ is the number of elements in the vector. The expected length of the noise intensity vector is given by:
\begin{equation}
    E\left(||\delta\mathbf{I}||\right)=\sqrt{E\left(\delta\mathbf{I}^\top\delta\mathbf{I}\right)}=\sqrt{E\left(\delta\mathbf{I}^\top U \delta\mathbf{I}\right)},
    \label{noisevector}
\end{equation}
with $U$ the unit matrix.
In Eq.~(\ref{noisevector}) the expected value of the quadratic form of a random vector ($\mathbf{X}$) can be recognised, this form can be rewritten in the following way:
\begin{equation}
E\left(\mathbf{X}^\top A \mathbf{X}\right)=E\left(\mathbf{X}^\top\right)AE\left(\mathbf{X}\right)+\mathcal{T}\left(AK_{\mathbf{XX}}\right),
\end{equation}
where $\mathcal{T}$ is trace of a matrix and $K_{\mathbf{XX}}$ is the auto-covariance matrix of the random vector. Using this property Eq.~(\ref{noisevector}) can rewritten as:
\begin{equation}
    E\left(||\delta\mathbf{I}||\right)=\sqrt{E\left(\delta\mathbf{I}^\top\right)UE\left(\delta\mathbf{I}\right)+\mathcal{T}\left(UK_{\delta\mathbf{I}\delta\mathbf{I}}\right)}.  
\end{equation}
Since the mean of each element in the random vector is zero, the first term in the square root vanishes and diagonal of the auto-covariance matrix will be equal to the variance of each element of the vector, resulting in:
\begin{equation}
    E\left(||\delta\mathbf{I}||\right)=\sqrt{\sum_{n=1}^N\sigma_n^2}=\sqrt{\sum_{n=1}^N\frac{1}{\mathrm{SNR}^2}}=\frac{\sqrt{N}}{\mathrm{SNR}},
\end{equation}
where $n$ is the element number of the vector. Now the expected length of both the signal and noise vector of the measured intensities is determined, the condition number of the reconstruction matrix ($k(A)$) as defined in Eq.~(\ref{cond}) can be use to determine the noise of the reconstructed intensities ($I_{pol}^\lambda$), resulting in:
 \begin{equation}
 \frac{||\delta \mathbf{I}_{pol}^\lambda||}{||\mathbf{I}_{pol}^\lambda||}=k(A)\frac{||\delta\mathbf{I}||}{||\mathbf{I_s}||}=\frac{k(A)}{SNR},
 \end{equation}
where $||\delta \mathbf{I}_{pol}^\lambda||$ is the expected length of the noise vector of the reconstructed intensities and $||\mathbf{I}_{pol}^\lambda||$ is the length of the actual reconstructed intensities. Since Gaussian function remain Gaussian under linear transformations, the noise in the reconstructed intensities will still be Gaussian. If now it assumed that the noise is spread equally over all reconstructed intensities, a reconstructed intensity for a single polarization and wavelength is given by:
\begin{equation}
    \tilde{I}_{pol}^\lambda=I_{pol}^\lambda+\mathcal{N}\left(0,\left(\frac{k(A)}{\mathrm{SNR}}\right)^2\right),
\end{equation}
with the tilde differentiating the true value $I_{pol}^\lambda$ from the obtain value with noise $\tilde{I}_{pol}^\lambda$. When using the definition of the Stokes parameters Eqs.~(\ref{S0}-\ref{S2}) and the summing properties of the Gaussian functions:
\begin{equation}
    \mathcal{N}\left(\mu_1+\mu2,\sigma_1^2+\sigma_2^2\right)=\mathcal{N}\left(\mu_1,\sigma_1^2\right)+\mathcal{N}\left(\mu_2,\sigma_2^2\right),
\end{equation}
it can be found that the second and the third estimated Stokes parameters are:
\begin{align}
    \tilde{S_1}&=I_x-I_y+\mathcal{N}\left(0,\left(\frac{\sqrt{2}k(A)}{\mathrm{SNR}}\right)^2\right)\\
    \tilde{S}_2&=I_{135^\circ}-I_{135^\circ}+\mathcal{N}\left(0,\left(\frac{\sqrt{2}k(A)}{\mathrm{SNR}}\right)^2\right).
\end{align}
The zeroth Stokes parameter is measured twice according to both definition in Eq.~(\ref{S0}) and is averaged between the two measurements, so its given by:
\begin{align}
\tilde{S}_0&=\frac{I_x+I_y+\mathcal{N}\left(0,\left(\frac{\sqrt{2}k(A)}{\mathrm{SNR}}\right)^2\right)}{2}+\frac{I_{135^\circ}+I_{135^\circ}+\mathcal{N}\left(0,\left(\frac{\sqrt{2}k(A)}{\mathrm{SNR}}\right)^2\right)}{2}\\
&=\frac{I_x+I_y}{2}+\frac{I_{135^\circ}+I_{135^\circ}}{2}+\mathcal{N}\left(0,\left(\frac{k(A)}{\mathrm{SNR}}\right)^2\right).
\end{align}
From these results Eqs.~(\ref{S0est}-\ref{S2est}) are easily obtained.

\section*{Funding}
Part of the activities have been supported by the TNO internal program: SMO 2018/19 Space Scientific Instruments. 
L.P.S. was funded by TNO (Netherlands Organisation for Applied Scientific Research) under the TU/e (Eindhoven University of Technology) program: 10022593 IMPULS II: Metrology 4 3D nano. J.B. has received funding from European Union's Horizon 2020 research and innovation programme under the Marie Sklodowska-Curie grant agreement No. 675745. 
F.S. acknowledges support by German Ministry of Education and Research (FKZ~03ZZ0434, FKZ~13N14877). F.E. and H.K. acknowledge support by German Ministry of Education and Research (ID~13XP5053A), and the support by the Max Planck School of Photonics.

\section*{Acknowledgments}
The authors are grateful to Dennis Arslan and Isabelle Staude for access to spectroscopy setup and related technical assistance, Michael Steinert for SEM images, and Pallabi Paul and Adriana Szeghalmi for ALD of SiO$_2$ cladding. The authors acknowledge Daniel Voigt, Holger Schmidt, Thomas K\"{a}sebier and J\"{o}rg Fuchs for technical assistance in nanostructuring of Si, Zuzanna Deutschman for initial simulations of the polarization-sensitive FP cavities, and Tiberiu Ceccotti for extensive discussions and helpful suggestions. 

\section*{Disclosures}
The authors declare no conflicts of interest related to this article.

\bibliography{My_Collection}

%%%%%%%%%%%%%%%%%%%%%%%%%%%%%%%%%%%%%%%%%%%%%%%%%%%%%%%%%%%

\end{document}